\documentclass[12pt]{article}

\makeatletter

\RequirePackage{cite}
\RequirePackage{times}
\RequirePackage{fullpage}
\RequirePackage{ifthen}

\renewcommand\@biblabel[1]{#1.}

\def\@cite#1#2{$^{\mbox{\scriptsize #1\if@tempswa , #2\fi}}$}

\newcommand{\spacing}[1]{\renewcommand{\baselinestretch}{#1}\large\normalsize}
\spacing{2}

\def\@maketitle{%
  \newpage\spacing{1}\setlength{\parskip}{12pt}%
    {\Large\bfseries\noindent\sloppy \textrm{\@title} \par}%
    {\noindent\sloppy \@author}%
}

\newenvironment{affiliations}{%
    \setcounter{enumi}{1}%
    \setlength{\parindent}{0in}%
    \slshape\sloppy%
    \begin{list}{\upshape$^{\arabic{enumi}}$}{%
        \usecounter{enumi}%
        \setlength{\leftmargin}{0in}%
        \setlength{\topsep}{0in}%
        \setlength{\labelsep}{0in}%
        \setlength{\labelwidth}{0in}%
        \setlength{\listparindent}{0in}%
        \setlength{\itemsep}{0ex}%
        \setlength{\parsep}{0in}%
        }
    }{\end{list}\par\vspace{12pt}}

\renewenvironment{abstract}{%
    \setlength{\parindent}{0in}%
    \setlength{\parskip}{0in}%
    \bfseries%
    }{\par\vspace{-6pt}}

\renewcommand{\section}{\@startsection {section}{1}{0pt}%
    {-6pt}{1pt}%
    {\bfseries}%
    }
\renewcommand{\subsection}{\@startsection {subsection}{2}{0pt}%
    {-0pt}{-0.5em}%
    {\bfseries}*%
    }

\newcommand*{\addendumlabel}[1]{\textbf{#1}\hspace{1em}}

\renewenvironment{figure}{\let\caption\NAT@figcaption}{}
    
\newcommand{\NAT@figcaption}[2][]{{%
    \refstepcounter{figure}
    \ifthenelse{\value{figure}=1}{
    }{
    }
    \noindent{\textbf{Fig.~\arabic{figure}~$|$~}}#2}
    }

\newcommand{\NAT@extfigcaption}[2][]{{%
    \refstepcounter{figure}
    \ifthenelse{\value{figure}=1}{
    }{
    }
    \noindent{\textbf{Extended Data Fig.~\arabic{figure}~$|$~}}#2}
    }

\newcommand{\NAT@exttabcaption}[2][]{{%
    \refstepcounter{table}
    \ifthenelse{\value{table}=1}{
    }{
    }
    \noindent{\sffamily\textbf{Extended Data Table~\arabic{table}~$|$~}}#2}
    }

\newcommand{\supplementary}{%
\setcounter{table}{0}
\setcounter{figure}{0}
\renewenvironment{figure}{\let\caption\NAT@suppfigcaption}{}
\renewenvironment{table}{\let\caption\NAT@supptabcaption}{}
}

\newcommand{\NAT@suppfigcaption}[2][]{{%
    \refstepcounter{figure}
    \ifthenelse{\value{figure}=1}{
    }{
    }
    \noindent{\sffamily\textbf{Supplementary Fig.~\arabic{figure}~$|$~}}#2}
    }

\newcommand{\NAT@supptabcaption}[2][]{{%
    \refstepcounter{table}
    \ifthenelse{\value{table}=1}{
    }{
    }
    \noindent{\sffamily\textbf{Supplementary Table~\arabic{table}~$|$~}}#2}
    }

\renewenvironment{table}{\@float{table}[p]\sffamily}{\end@float}

\newcommand{\extended}{%
\setcounter{table}{0}
\setcounter{figure}{0}
\renewenvironment{figure}{\let\caption\NAT@extfigcaption}{}
\renewenvironment{table}{\let\caption\NAT@exttabcaption}{}
}

\newcommand{\NAT@ignore}[2][]{}

\makeatother

\newcommand{\kepler}[0]{Kepler}
\newcommand{\galaxia}[0]{\emph{Galaxia}}

\newcommand{\gaia}[0]{\emph{Gaia}}
\newcommand{\Teff}[0]{$T_{\text{eff}}$}

\newcommand{\Dnu}[0]{$\Delta\nu$}

\newcommand{\numax}[0]{$\nu_{\rm max}$}

\newcommand{\DP}[1]{$\Delta P_{#1}$}

\newcommand{\muHz}[0]{$\mu{\rm Hz}$}

\newcommand{\Msun}[0]{M$_{\odot}$}
\newcommand{\Rsun}[0]{R$_{\odot}$}
\newcommand{\Lsun}[0]{L$_{\odot}$}

\usepackage{CJKutf8}
\newcommand{\CNnames}[1]{{\begin{CJK}{UTF8}{gbsn}~(#1)~\end{CJK}}}

\let\citep\cite
\let\citet\cite

\usepackage{graphicx}	
\usepackage{amsmath}	
\usepackage{amssymb}	
\usepackage{xcolor}
\usepackage[normalem]{ulem}
\usepackage{tabularx} 
\usepackage{booktabs}
\usepackage{multirow}

\newcommand{\apj}{Astrophys. J.}
\newcommand{\apjl}{Astrophys. J.}
\newcommand{\apjs}{Astrophys. J. Suppl. Ser.}
\newcommand{\aap}{Astron. Astrophys.}
\newcommand{\aapr}{Astron. Astrophys. Rev.}

\newcommand{\araa}{Ann. Rev. Astron. Astrophys.}

\newcommand{\mnras}{Mon. Not. R. Astron. Soc.}
\newcommand{\aj}{Astron. J.}
\newcommand{\nat}{Nature}
\newcommand{\apss}{Astrophys. Space Sci.}
\newcommand{\pasp}{Pub. Astron. Soc. Pac.}

\spacing{1}
\newcommand{\firsttitle}[1]{\noindent\textbf{\large #1}\medskip}
\newcommand{\secondtitle}[1]{\textbf{#1}}

\newcommand{\SIfA}{\small\textbf{\sffamily{1}}}
\newcommand{\SAC}{\small\textbf{\sffamily{2}}}
\newcommand{\UNSW}{\small\textbf{\sffamily{3}}}
\newcommand{\IfA}{\small\textbf{\sffamily{4}}}
\newcommand{\STSI}{\small\textbf{\sffamily{5}}}
\newcommand{\BNU}{\small\textbf{\sffamily{6}}}
\newcommand{\AMNH}{\small\textbf{\sffamily{7}}}
\newcommand{\IRAP}{\small\textbf{\sffamily{8}}}
\newcommand{\UNSWDSH}{\small\textbf{\sffamily{9}}}
\newcommand{\NAOC}{\small\textbf{\sffamily{10}}}

\newcommand{\Nofunderluminous}{7}
\newcommand{\Nofundermassive}{32}
\newcommand{\Nofheb}{7538}

\newcommand{\showfigures}[1]{#1\\}
\usepackage{academicons}
\usepackage{hyperref}
\usepackage[notransparent]{svg}
\newcommand{\orcid}[1]{\href{https://orcid.org/#1}{\textsuperscript{\includegraphics[width=10pt]{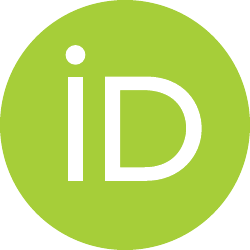}}}}

\newcommand{\rev}[1]{#1}

\title{Discovery of post-mass-transfer helium-burning red giants using asteroseismology}

\author{\small\sffamily\textbf{Yaguang~Li\CNnames{李亚光}\thanks{e-mail: yaguang.li@sydney.edu.au}\orcid{0000-0003-3020-4437}$^{\SIfA,\SAC}$, 
Timothy~R.~Bedding\thanks{e-mail: tim.bedding@sydney.edu.au}\orcid{0000-0001-5222-4661}$^{\SIfA,\SAC}$,
Simon~J.~Murphy\orcid{0000-0002-5648-3107}$^{\SIfA,\SAC}$,
Dennis~Stello\orcid{0000-0002-4879-3519}$^{\UNSW,\SAC}$,\\
Yifan~Chen\CNnames{陈逸凡}\orcid{0000-0001-6020-0483}$^{\SIfA}$,
Daniel~Huber\orcid{0000-0001-8832-4488}$^{\IfA}$,
Meridith~Joyce\orcid{0000-0002-8717-127X}$^{\STSI}$,
Dion~Marks$^{\SIfA}$,\\
Xianfei~Zhang\CNnames{张先飞}\orcid{0000-0002-3672-2166}$^{\BNU}$,
Shaolan~Bi\CNnames{毕少兰}\orcid{0000-0002-7642-7583}$^{\BNU}$,
Isabel~L.~Colman\orcid{0000-0001-8196-516X}$^{\AMNH}$,\\
Michael~R.~Hayden\orcid{0000-0001-7294-9766}$^{\SIfA}$,
Daniel~R.~Hey\orcid{0000-0003-3244-5357}$^{\SIfA,\SAC}$,
Gang~Li\CNnames{李刚}\orcid{0000-0001-9313-251X}$^{\IRAP}$,
Benjamin~T.~Montet\orcid{0000-0001-7516-8308}$^{\UNSW,\UNSWDSH}$,\\
Sanjib~Sharma\orcid{0000-0002-0920-809X}$^{\SIfA}$
and Yaqian~Wu\CNnames{武雅倩}\orcid{0000-0002-8337-4117}$^{\NAOC}$\\
}}

\begin{document}

\maketitle
\begin{affiliations}
\small
\medskip
\item Sydney Institute for Astronomy (SIfA), School of Physics, University of Sydney, Camperdown, NSW 2006, Australia.
\item Stellar Astrophysics Centre, Department of Physics and Astronomy, Aarhus University, Aarhus, Denmark.
\item School of Physics, University of New South Wales, Kensington, New South Wales, Australia.
\item Institute for Astronomy, University of Hawai`i, 2680 Woodlawn Drive, Honolulu, HI 96822, USA.
\item Space Telescope Science Institute, 3700 San Martin Dr, Baltimore, MD 21218, USA.
\item Department of Astronomy, Beijing Normal University, Haidian District, Beijng 100875, China.
\item Department of Astrophysics, American Museum of Natural History, 200 Central Park West, Manhattan, NY, USA.
\item IRAP, Université de Toulouse, CNRS, CNES, UPS, Toulouse, France.
\item UNSW Data Science Hub, University of New South Wales, Sydney, NSW 2052, Australia.
\item Key Laboratory of Optical Astronomy, National Astronomical Observatories, Chinese Academy of Sciences, A20 Datun Road, Chaoyang District, Beijing 100101, China.
\end{affiliations}


\begin{abstract}
A star expands to become a red giant when it has fused all the hydrogen in its core into helium. If the star is in a binary system, its envelope can overflow onto its companion or be ejected into space, leaving a hot core and potentially forming a subdwarf-B star\citep{heber-2016-sdb-review,byrne++2021-population-sdb,lynas-gray-2021-sdb-review}.
However, most red giants that have partially transferred envelopes in this way remain cool on the surface and are almost indistinguishable from those that have not. 
Among $\sim$7000 helium-burning red giants observed by NASA's \kepler{} mission, we use asteroseismology to identify two classes of stars that must have undergone dramatic mass loss, presumably due to stripping in binary interactions.
The first class comprises about \Nofunderluminous{} underluminous stars with smaller helium-burning cores than their single-star counterparts.
Theoretical models show that these small cores imply the stars had much larger masses when ascending the red giant branch.
The second class consists of \Nofundermassive{} red giants with masses down to 0.5 \Msun{}, whose implied ages would exceed the age of the universe had no mass loss occurred.
The numbers are consistent with binary  statistics, and
our results open up new possibilities to study the evolution of post-mass-transfer binary systems.
\end{abstract}
\bigskip

Mass loss in red giant stars remains one of the major uncertainties in stellar physics. 
A hydrogen-shell-burning red-giant-branch (RGB) star will reach its maximum luminosity at the tip of the RGB, where substantial mass loss occurs\citep{reimers-1975-wind,schroder+2005-wind}. It then starts the core helium-burning (CHeB) phase at a much lower luminosity.
Recent studies suggest that the accumulated mass loss driven by pulsation and radiation on the RGB can reduce the stellar mass by up to 0.1 \Msun{}, based on asteroseismic observations of field stars\citep{yuj++2021-rg-mass-loss,miglio++2021-age-kepler} and open clusters\citep{miglio++2012-mass-loss-ngc6791-ngc6819,stello++2016-m67,handberg++2017-ngc6819}.
In contrast, globular clusters tend to suggest a loss in mass of about $0.2$~\Msun{} on the RGB based on the morphology of the horizontal-branch on the H--R diagram\citep{mcdonald+2015-rgb-mass-loss,lebzelter+2011-ngc362-ngc2808,salaris++2016-47-Tuc},
although the accuracy of photometric masses is still being debated\citep{an++2019-photometric-mass}.
Even greater changes in mass can occur during binary interactions, via stable Roche lobe overflow, common envelope ejection, or merging\citep{hanzw++20-binary-review}. 

The fate of an RGB star in a binary system can vary markedly, depending on the system's dynamical properties and hence the mass transfer rate.  
If the star loses its entire hydrogen-rich envelope before reaching the RGB tip, it leaves a bare non-burning helium core, forming a low-mass white dwarf\citep{bergeron++1992-white-dwarf,liebert++2005-white-dwarf,brown++2013-elm-5} ($M/$\Msun{}$<0.5$; $M$ is the stellar mass).
On the other hand, a stripped CHeB red giant could form a hot subluminous star of spectral type B (sdB) on the extreme horizontal branch\citep{byrne++2021-population-sdb,lynas-gray-2021-sdb-review,hanzw++2002-sdb-origin,huhl++2008-sdb-formation}.
Indeed, most sdB stars are found to be in binary systems with short periods\citep{maxted++2001-binary-ehb,napiwotzki++2004-rv-sdb,copperwheat++2011-rv-sdb}. 
Some stripped core-helium-burning stars
are found in binary systems with a Be star (B star with a circumstellar disc) as the companion\citep{el-badry+2021-hr6819-lb1,shenar++2020-lb1,irrgang++2020-lb1},
suggesting a mass-transfer history.
However, there has been little success in finding CHeB red giant stars that have only partially transferred their envelopes, except in a few open clusters where an anomaly in stellar mass is more easily identified\citet{handberg++2017-ngc6819,brogaard++2021-ngc6791}.

To find these post-mass-transfer CHeB stars among the red giants observed by \kepler{}, we used asteroseismology to derive stellar parameters and evolutionary phases (see Methods).
According to the asteroseismic scaling relations\citep{ulrich-1986-age,brown++1991-dection-procyon-scaling-relation,kjeldsen+1995-scaling-relations}, the so-called large frequency separation scales with the mean density, $\Delta\nu \propto M^{1/2}R^{-3/2}$ ($R$ is the stellar radius), and the frequency of maximum oscillation power is proportional to the surface properties, $\nu_{\rm max} \propto g/\sqrt{T_{\rm eff}} \propto MR^{-2}T_{\rm eff}^{-1/2}$ ($g$ is the surface gravity; \Teff{}, the effective temperature). 
These two relations give stellar masses and radii to remarkable precision\citep{liyg++2021-sc-intrinsic-scatter}. 
In addition, the non-radial oscillation modes of red giants (spherical degree $l\geq1$) are mixed modes, which result from coupling between gravity (g) waves in the core and acoustic pressure (p) waves in the envelope\citep{aizenman++1977-avoided-crossing,jcd++1995-eta-boo,deheuvels++2010-hd49385-corot,benomar++2013-avoided-crossing-sg-kepler}. 
The period spacing of the $l=1$ modes, \DP{}, is a reliable indicator to distinguish CHeB from RGB stars\citep{dupret-2009-theoretical-amp-lw-nonradial-solarlike-rg,bedding++2011-distinguish-rc-rgb}.

\begin{figure}
\centering
\showfigures{\includegraphics[width=\linewidth]{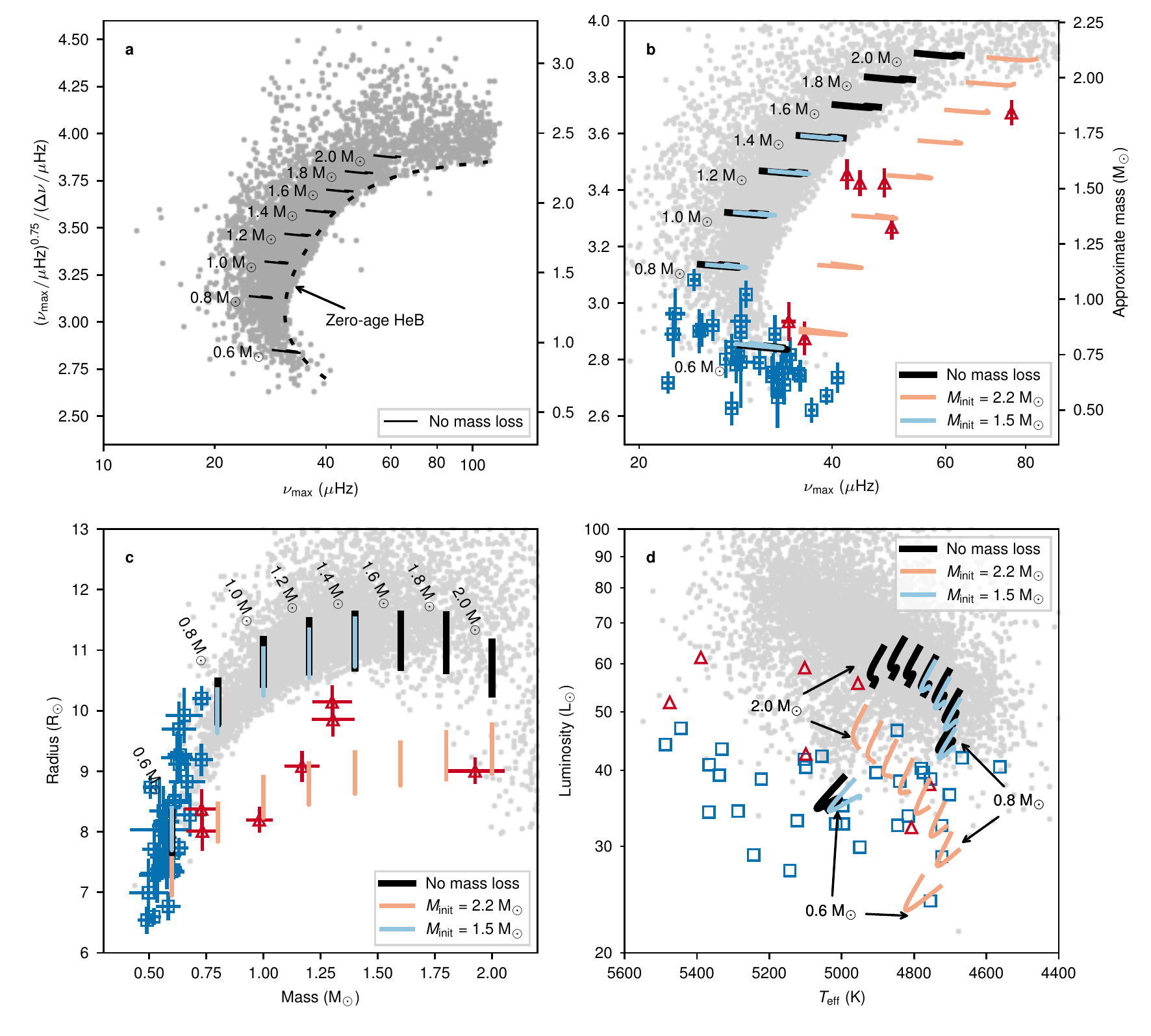}}
\caption{Fundamental parameters of CHeB stars in the \kepler{} red giant sample. a) and b) the seismic quantity $\nu_{\rm max}^{0.75}/\Delta\nu$ vs. \numax{}.  c) the mass--radius diagram. d) the H--R diagram. The underluminous stars are shown in red triangles, the very low-mass stars in blue squares, and the rest of the CHeB stars in grey points. 
The final masses of stellar evolutionary tracks are marked by numbers in \Msun{}. The tracks were calculated without mass loss (black lines), with mass loss from initial mass of 1.5 \Msun{} (light blue lines), and with mass loss from initial mass of 2.2 \Msun{} (orange lines). The underluminous stars were identified as stars lying $>1\sigma$ to the ZAHeB edge (panel a).
\rev{Data points are median values. Error bars on the underluminous and very low-mass stars in panel b and c show 1$\sigma$ uncertainties. 
Other data points are not shown with error bars to improve clarity.} \label{fig:nike}}
\end{figure}

Fig.~\ref{fig:nike} shows the parameters for \Nofheb{} CHeB stars in our sample. 
Since both \Dnu{} and \numax{} depend on radius, we examine the quantity $\nu_{\rm max}^{0.75}/\Delta\nu$ in Fig.~\ref{fig:nike}a (both \numax{} and \Dnu{} are in \muHz{}).
According to the scaling relations, $\nu_{\rm max}^{0.75}/\Delta\nu$ is proportional to $ M^{0.25}T_{\rm eff}^{-0.375}$ and is approximately independent of radius\citep{huber++2010-800-rg-kepler,bedding++2011-distinguish-rc-rgb,yuj++2018-16000-rg}. 
The most notable feature in Fig.~\ref{fig:nike}a is a hook-like structure with almost all stars sitting on the one side of a well-defined edge, which corresponds to the zero-age helium-burning (ZAHeB) phase\citep{liyg++2021-sc-intrinsic-scatter}. 
This ZAHeB edge is very sharp because almost all ZAHeB stars with $M$ $\lesssim1.8$ \Msun{} share a common helium core mass of $\sim0.5$ \Msun{}\citep{sweigart-1990-rgb, montalban++2013-overshoot}, which was supported by electron degeneracy on the RGB (see Extended Data Fig.~\ref{fig:galaxia}). 
We calculated CHeB stellar models from $0.6$ \Msun{} to $2.0$ \Msun{} with solar metallicity and assuming single-star evolution (see Methods), shown by the black lines. 
Overall, these models are consistent with the majority of the CHeB population, especially considering the single metallicity and the neglection of convective overshoot\citep{girardi-2016-rc-review}.

The group of stars in Fig.~\ref{fig:nike}a that lie to the right of the ZAHeB edge cannot be explained by single-star evolution. 
They are smaller in radius, and hence lower in luminosity, than the main CHeB population with the same masses, implying that a smaller core is supplying their energy. 
We refer to them as underluminous stars (see Extended Data Figs.~\ref{fig:spectra} and~\ref{fig:echelle} for an example).

Fig.~\ref{fig:nike}a also reveals a set of stars with masses down to $0.5$ \Msun{}. Modelling of the individual frequencies confirms the low mass (see Methods and Extended Data Fig.~\ref{fig:sapphire}).
The age of the universe, 13.8\,Gyr\citep{planck++2015-primordial-helium}, puts a lower limit on the mass of a red giant without mass loss to be approximately $0.8-1.0$~\Msun{}. 
Specifically, in Fig.~\ref{fig:mass-feh}, we show this lower limit on mass as a function of [M/H], determined by theoretical models (see Methods).
Since stellar winds driven by radiation and pulsation can only remove up to $0.2$ \Msun{} on the RGB, those stars below the threshold must have undergone much more extreme mass loss (see Extended Data Fig.~\ref{fig:galaxia}). 
We refer to them as very low-mass stars (see Extended Data Figs.~\ref{fig:spectra} and~\ref{fig:echelle} for an example).

Fig.~\ref{fig:nike}b highlights the underluminous stars (red triangles) and the very low-mass stars (blue squares).
We show the sample on the mass--radius diagram in Fig.~\ref{fig:nike}c, calculated from \Teff{}, \Dnu{} and \numax{} using the scaling relations, and on the H--R diagram in Fig.~\ref{fig:nike}d\citep{mosser++2012-amplitude-rg,elsworth++2019-evol-stages,gaulme++20-active-rg,liyg++2021-sc-intrinsic-scatter}.
The ZAHeB edge is still evident in the mass--radius plane (Fig.~\ref{fig:nike}c), though less sharp, due to observational uncertainties in \Teff{}.
The ZAHeB edge is not visible in the luminosity--\Teff{} plane (Fig.~\ref{fig:nike}d), presumably because \Teff{} depends strongly on both mass and metallicity on the red giant branch.
This reasoning is supported by the fact that the solar-metallicity evolutionary models in Fig.~\ref{fig:nike}d are unable to cover the whole observed \Teff{} range.


To understand the locations of the underluminous and very low-mass stars, we calculated stellar evolutionary models with various amounts of mass loss due to binary stripping (see Methods). They are shown by the tracks in Fig.~\ref{fig:nike}b--d. 
Firstly, the models with a progenitor mass of $2.2$ \Msun{} that lose different fractions of their outer envelopes (shown in orange lines) lie to the right of the ``hook'' formed by the CHeB population, in the same location as the underluminous stars. 
This confirms that the underluminous stars were originally more massive on the RGB ($1.8<M/{\rm M_{\odot}}<3.6$), 
where the central temperature rose quickly and the core started to collapse once reaching the Schn{\"o}berg-Chandrasekhar limit\citep{schonberg-1942-sc-limit}. 
This limit does not apply to the lower-mass RGB stars ($M/{\rm M_{\odot}}\lesssim 1.8$) because their dense cores are electron degenerate.
Hence, at the end of the RGB, the higher-mass stars ($1.8<M/{\rm M_{\odot}}<3.6$) initiated helium burning earlier and formed smaller helium cores than lower-mass stars ($M/{\rm M_{\odot}}\lesssim 1.8$).

Secondly, we show models with a progenitor mass of $1.5$ \Msun{} that lost different amounts of mass due to binary stripping (light blue lines). Their locations are almost the same as those without mass loss (black lines). 
This is because, after losing part of their envelope, their structure in the CHeB stage is essentially identical to a star that began its life with that lower mass.
It is therefore impossible to decipher how much mass a star has lost based on its current \Dnu{}, \numax{}, luminosity and \Teff{} if it was born with an initial mass below $1.8$ \Msun{}. 
However, the $0.6$-\Msun{} models without mass loss are older than the universe, 
while the mass-loss models produce realistic ages for the very low-mass stars.

Based on known binary distributions, we can calculate the number of \kepler{} red giants expected to undergo mass loss after filling their Roche lobes (see Methods). 
The resulting fraction of underluminous stars with progenitor masses between $1.8$ and $3.6$ \Msun{} is predicted to be $0.13\%$. That is consistent with our observation, $0.09\%\pm0.04\%$. 
The predicted fraction of post-mass-transfer stars with progenitor masses below $1.8$ \Msun{} is $2.01\%$ --- more than the $0.48\%\pm0.09\%$ very low-mass CHeB stars we observed.
This is to be expected, because post-mass-transfer CHeB stars with $1.0<M/$\Msun{}\,$<1.8$ hide in the overall CHeB population (grey points in Fig.~\ref{fig:nike}).

Although the post-mass-transfer stars that now appear as regular CHeB stars are difficult to identify, future studies of chemical abundances may provide clues on mass transfer.
One example is lithium (Li), an element that cannot survive in a high-temperature environment. 
In red giants, the expansion of the convective envelope dilutes Li on the surface by bringing Li-deficient layers from below. 
Hence, Li enhancement [$A({\rm Li})=\log_{10}\left(n({\rm Li})/n({\rm H})) \right)>1.5$ dex, where $n({\rm x})$ is the number density of atom ${\rm x}$] in red giant stars is unusual\citep{kumar++2020-li-rgs,deepak+2021-lithium-seismology,martell-2021-galah-lithium,yanhl++2021-li-rgs}.
Among the underluminous stars (red triangles in Fig.~\ref{fig:nike}), KIC 5000307 shows an unusually high abundance of Li\citep{silvaaguirre++2014-kic5000307-li}, with $A({\rm Li})=2.8$. 
Our result showing this star has experienced dramatic mass loss seems to suggest binarity as a Li production channel\citep{casey++2019-li-tidal,zhangxf++2020-li-merger}.


Another group of interest are the $\alpha$-process elements, which trace the stellar populations in the Galaxy.
In particular, the $\alpha$-rich population is characterised by its old age\citep{jbh+2016-galaxy}, mainly consisting of low-mass stars, as shown in Fig~\ref{fig:alpha}a by the [$\alpha$/M]$>0.15$ sample. 
However, this population also contains the so-called young-alpha-rich stars\citep{chiappini++2015-young-alpha-rich,martig++2015-young-alpha-rich}, which appear old chemically but have large masses. They are suggested to be in wide binary systems and have been recipients of mass transfer\citep{jofre++2016-young-alpha-rich,yong++2016-young-alpha-rich,hekker+2019-young-alpha-rich,zhangm++2021-young-alpha-rich}.
Studies of their companions may be a way to find more stripped CHeB stars.

The elemental abundances of individual stars depend on age and metallicity [M/H]\citep{sharma++2020-galah-abundance-age,hayden++2020-galah-chemical-clocks}, 
which means stars within a specific [$\alpha$/M] range share a common age distribution. 
Fig.~\ref{fig:alpha}b shows the mass distributions of RGB and CHeB stars with $[\alpha$/M]$<0.03$. 
Almost all the RGB stars are more massive than 1 \Msun{}. Considering a maximum mass loss of 0.2 \Msun{} on the RGB through radiation and pulsation, the $M<0.8$ \Msun{} CHeB stars with $[\alpha$/M]$<0.03$ must have transferred mass by other means. 
Chemical abundances allow us to identify more of these stars in this way.

Our discovery of the post-mass-transfer CHeB stars follows recent identifications of mergers on the RGB\citep{rui+2021-rgb-merger,deheuvels++2021-rgb-merger} and demonstrates asteroseismology as a new way to find interesting binary systems in the red giant population.
Expanding the current sample to brighter stars from the K2 and TESS missions will enable spectroscopic or astrometric measurements to solve the binary orbits, allowing detailed characterisation of the systems and a better understanding of the mass transfer channel\citep{hanzw++2002-sdb-origin}. 
This is critical to investigate whether some of these stars are still undergoing mass loss and whether they will ultimately become sdB stars.
Asteroseismology also opens up other possibilities, since by modelling individual frequencies we can derive accurate masses and ages, thereby providing crucial constraints to the system's history.
Furthermore, analysing the rotational splitting of oscillation modes probes the core rotation and angular momentum transport surrounding these binary interactions, filling the gap between sdB stars and regular CHeB stars\citep{aerts++2019-am-review}.

\clearpage
\section*{Methods}

\bigskip

\noindent\secondtitle{Sample selection and stellar parameters} We used the asteroseismic red giant catalogue by Yu et al.\citep{yuj++2018-16000-rg}. 
This sample provides measurements of \Dnu{} and \numax{} from the SYD pipeline\citep{huber++2009-syd-pipeline}, compilations of \Teff{} and [M/H], and masses and radii derived using the asteroseismic scaling relations\citep{stello++2008-wire-kgiants,kallinger++2010-rg-corot-mass-radius,chaplin+2013-solar-like-review, hekker+2017-giant-review,basu+2020-giants-review,hekker-2020-scaling-review}.
\rev{It also compiles classifications of the evolutionary stage (RGB/CHeB) from previous work\citep{hon++2017-deep-learning-rc-rgb,kallinger++2012-epsp,stello-2013-classifcation-13000-rg-kepler,mosser++2014-mixed-mode-rg-window,vrard-2016-period-spacing-rg-2-automated-measures}.} The CHeB stars and the low-luminosity RGB stars ($\nu_{\rm max}>80$ \muHz{}) used in this work are all from this catalog.

\begin{figure}
\centering
\showfigures{\includegraphics[width=0.5\linewidth]{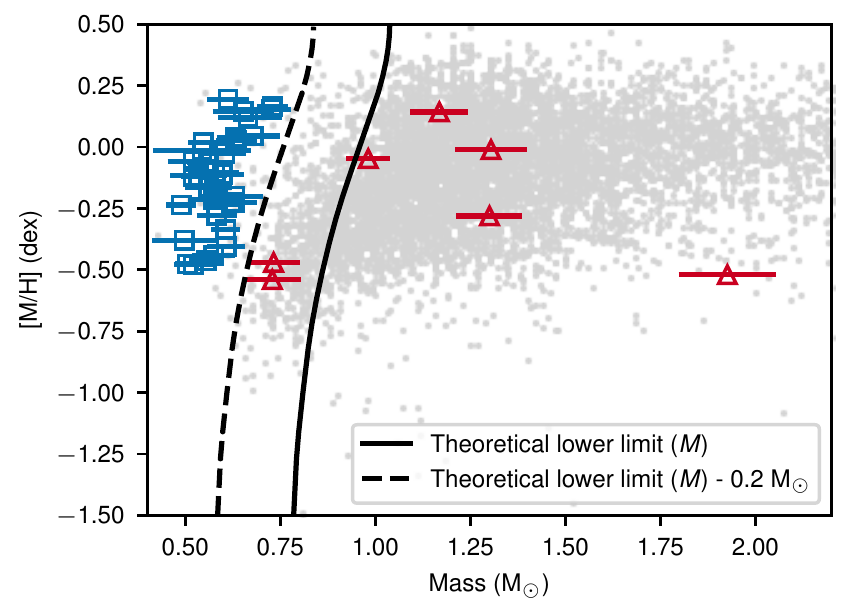}}
\caption{[M/H] vs. mass for CHeB stars in the \kepler{} red giant sample. The underluminous stars are shown as red triangles, the very low-mass stars as blue squares, and the rest of the CHeB stars as grey points. The theoretical lower limit on mass as a function of metallicity, as determined with stellar evolutionary models, is shown by the solid line, assuming no mass loss. Including a maximum possible mass loss of 0.2 \Msun{} driven by radiation and pulsation on the RGB\citep{miglio++2012-mass-loss-ngc6791-ngc6819,stello++2016-m67,handberg++2017-ngc6819,mcdonald++2011-omega-cen,lebzelter+2011-ngc362-ngc2808,mcdonald+2015-rgb-mass-loss,salaris++2016-47-Tuc} is shown by the dashed line. The very low-mass stars were identified as stars lying $>1\sigma$ to the left of the dashed line.
\rev{Data points are median values. Error bars on the underluminous and very low-mass stars show 1$\sigma$ uncertainties. 
Other data points are not shown with error bars to improve clarity.}\label{fig:mass-feh}}
\end{figure}

\begin{figure}
\centering
\showfigures{\includegraphics[width=\linewidth]{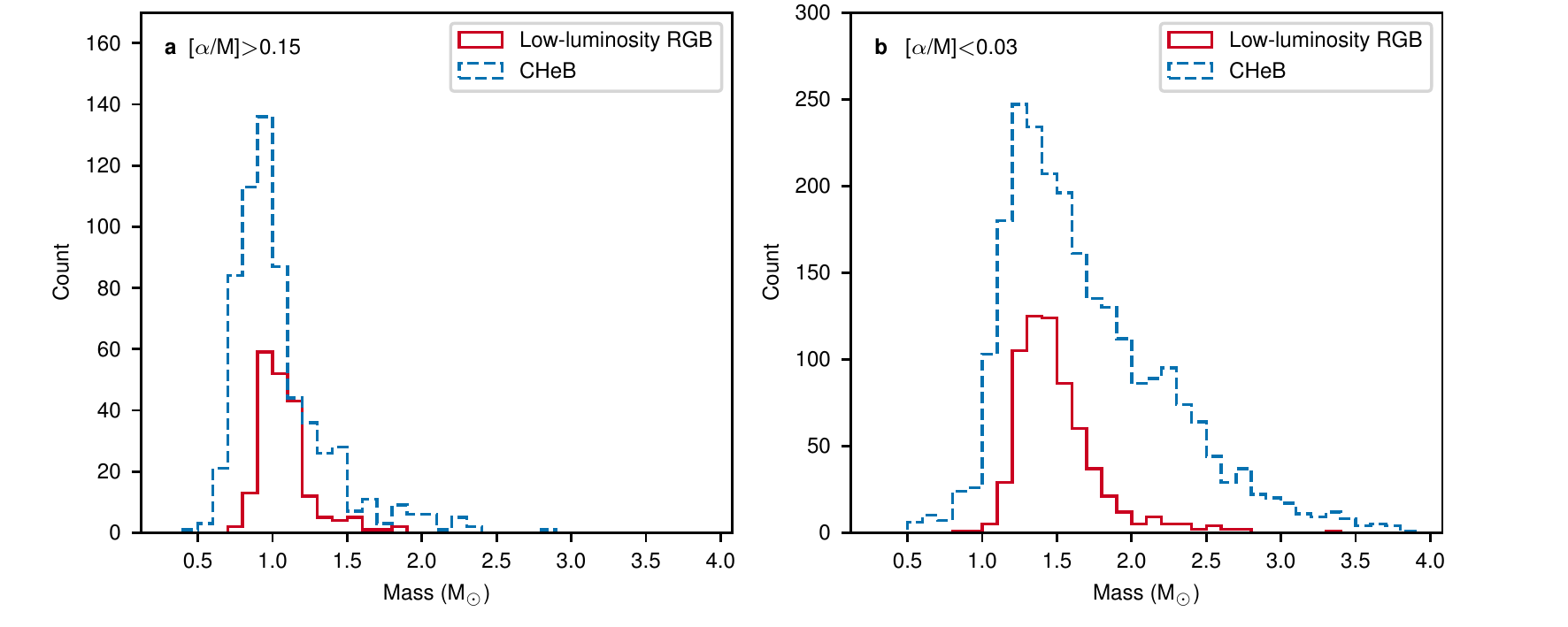}}
\caption{Mass distributions of low-luminosity RGB (\numax{}$>$80~\muHz{}) and CHeB stars in the \kepler{} red giant sample (see Methods). 
a) $\alpha$-rich population ([$\alpha$/M]$>0.15$). 
b) $\alpha$-poor population ([$\alpha$/M]$<0.03$).
\label{fig:alpha}}
\end{figure}

We cross-matched the sample with APOGEE DR17\citep{abdurrouf++2021-apogee-dr17} and LAMOST DR5\citep{xiang++2019-lamost-dr5} to obtain the elemental abundances ([M/H] and [$\alpha$/M]), replacing the values of [M/H] from Yu et al.\citep{yuj++2018-16000-rg} wherever possible. The elemental abundances are used in Figs.~\ref{fig:mass-feh} and~\ref{fig:alpha}.
We also obtained radial velocities from APOGEE DR17\citep{abdurrouf++2021-apogee-dr17} and the LAMOST medium resolution survey\citep{zhangb++2021-lamost-mrs-rv}.

We carefully re-measured the \numax{} values for the underluminous and low-mass stars in our sample using the pySYD pipeline\citep{chontos++2021-pysyd} and found good agreement with the catalogue values. 
In addition, we extracted the radial mode frequencies for these identified stars\citep{liyg++2020-kepler-36-subgiants} and used them to re-determine \rev{their average \Dnu{} by adopting the slope from fitting a straight line to the frequencies as a function of the radial orders\citep{white++2011-asteroseismic-diagrams-cd-epsilon-deltaP-models}.} 
This allowed us to measure \Dnu{} more accurately because it is less affected by one or more strong modes.
\rev{We calculated the correction factors for the \Dnu{} scaling relation according to Sharma et al.\citep{sharma++2016-population-rg-kepler}.}
The masses and radii (and associated uncertainties) were then re-determined using the re-derived \Dnu{} \rev{(with the associated correction factors) and the revised \Teff{}}, while keeping all the other parameters the same as described in Yu et al.\citep{yuj++2018-16000-rg}. 
We determined luminosities via the Stefan-Boltzmann law $L\propto R^2T_{\rm eff}^4$.
We also examined the classification results based on the period spacings \DP{} to confirm the evolutionary stages.
To do this, the power spectrum in period was sliced into segments at an equal width and vertically stacked to construct the so-called period \'{e}chelle diagram. By optimising the width, the period spacing, \DP{}, could be obtained, such that $l=1$ modes align in a ``zig-zag'' pattern\citep{bedding++2011-distinguish-rc-rgb}. 
We checked the period spacings directly (rather than fitting with functions\citep{vrard++2016-period-spacings}) because the period spacings of CHeB and RGB populations (at similar \numax{}) differ by at least a factor of $4$ and inspection of the \'{e}chelle diagram is sufficient to assign the class of evolutionary stages\citep{vrard++2016-period-spacings,mosser++2018-period-spacing-4-complete-description-mixed-mode-patterns}.
Extended Data Fig.~\ref{fig:spectra} shows the power spectra for three representative stars, including a regular CHeB star (panel a), an underluminous star (panel b), and a very low-mass star (panel c). The spectra show clear detections of $l=1$ modes. Extended Data Fig.~\ref{fig:echelle} shows the period \'{e}chelle diagrams. Their period spacings are about $300$ s, confirming them as CHeB stars.
The main characteristics that set the post-mass-transfer stars apart are their values of \numax{} and \Dnu{}, and therefore their masses and radii.



\noindent\secondtitle{Stellar evolutionary models}
We calculated the stellar evolutionary models shown in Fig.~\ref{fig:nike} with MESA (Modules for Experiments in Stellar Astrophysics; version 15140)\citep{paxton++2011-mesa,paxton++2013-mesa,paxton++2015-mesa,paxton++2018-mesa,paxton++2019-mesa} and GYRE (version 6.0.1)\citep{townsend+2013-gyre}. 
We used the Henyey formalism\citet{henyey++1965-mlt} of the mixing length theory to describe convection, with the mixing length parameter $\alpha_{\rm MLT}$ set to $2$. Prior work in this mass regime suggests fundamental parameters are not sensitive to the choice of $\alpha_{\rm MLT}$ at this observational precision\citep{murphy++2021-hd139614,molnar++2019-T-ursae-minoris}. 
We did not include convective overshoot.
We adopted the current solar photospheric abundance measured by Asplund et al.\citet{asplund++2009-solar-composition-review} as the metal mixture for our calculation: $X_{\odot}=0.7381$, $Y_{\odot}=0.2485$, $Z_{\odot}=0.0134$. 
The opacity tables were accordingly set based on the AGSS09 metal mixture.
We used nuclear reaction rates from the JINA REACLIB database\citep{cyburt++2010-jina-reaclib} and only considered a minimal set of elements adopted from \texttt{basic.net} in {\small MESA}. 
We adopted the grey model atmosphere with Eddington $T$--$\tau$ integration\citep{eddington-1926-star} as the surface boundary condition. 
MESA uses a equation of state blended from OPAL\citep{rogers+2002-opal-eos}, SCVH\citep{saumon++1995-eos}, PTEH\citep{pols++1995-eos}, HELM\citep{timmes+2000-eos}, and PC\citep{potekhin+2010-eos}.
MESA implements electron conduction opacities\citep{cassisi++2007-opacity} and radiative opacities from OPAL\citep{iglesias+1993-opal,iglesias+1996-opal}, except low-temperature data\citep{ferguson++2005-opacity} and the high-temperature Compton-scattering regime\citep{buchler+1976-opacity}.

We implemented an instant (compared to the evolutionary timescales) mass loss to model a quick binary stripping process. When the model evolved to the CHeB stage, mass loss from the surface was switched on (with the ``mass\_change'' option in MESA) at a rate of $2$ \Msun{}/Myr. Because the helium burning lasts about $100$ Myr for $M<2.2$ \Msun{} stars\citep{stello-2013-classifcation-13000-rg-kepler,girardi-2016-rc-review}, and the total amount of mass loss ranges from $0.2$ to $1.6$ \Msun{}, the implemented mass loss spanned less than $1\%$ of the total CHeB stage.
Mass loss was turned off once the desired final mass was reached, and the evolution was continued until the exhaustion of core helium. 
Using these settings, we calculated evolutionary models with initial masses $1.5$ and $2.2$ \Msun{} and final masses ranging from $0.6$ to $2.0$ \Msun{} in steps of 0.1 \Msun{}. We also computed models without any mass loss spanning this mass range, for comparison.

\begin{figure}
    \centering
    \showfigures{\includegraphics[width=\columnwidth]{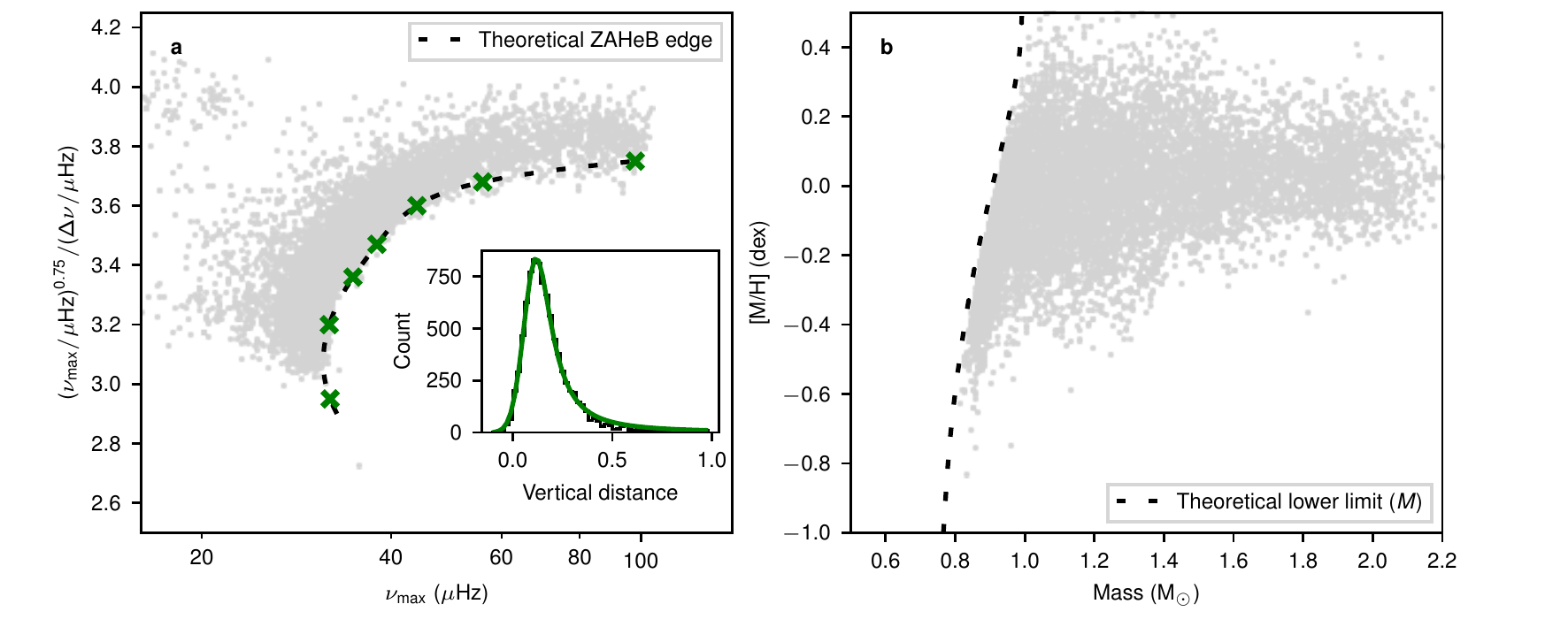}}
    \caption{\galaxia{} simulation of CHeB stars in the \kepler{} field. a): $\nu_{\rm max}^{0.75}/\Delta\nu$ vs. \numax{}. The ZAHeB edge (the black dashed line) is represented by a spline (defined by the crosses). The inset of a) shows the distribution of the vertical distances to the edge. The distribution is fitted by a half-Gaussian half-Lorentzian profile, shown by the green line. The standard deviation of the half-Gaussian profile represents the intrinsic broadening of the ZAHeB edge. b): the metallicity--mass diagram. The dashed line is the lowest mass a star can be without mass loss given a metallicity, determined with MIST models (see Methods).\label{fig:galaxia}} 
\end{figure}

\begin{figure}
\centering
\showfigures{\includegraphics[width=\linewidth]{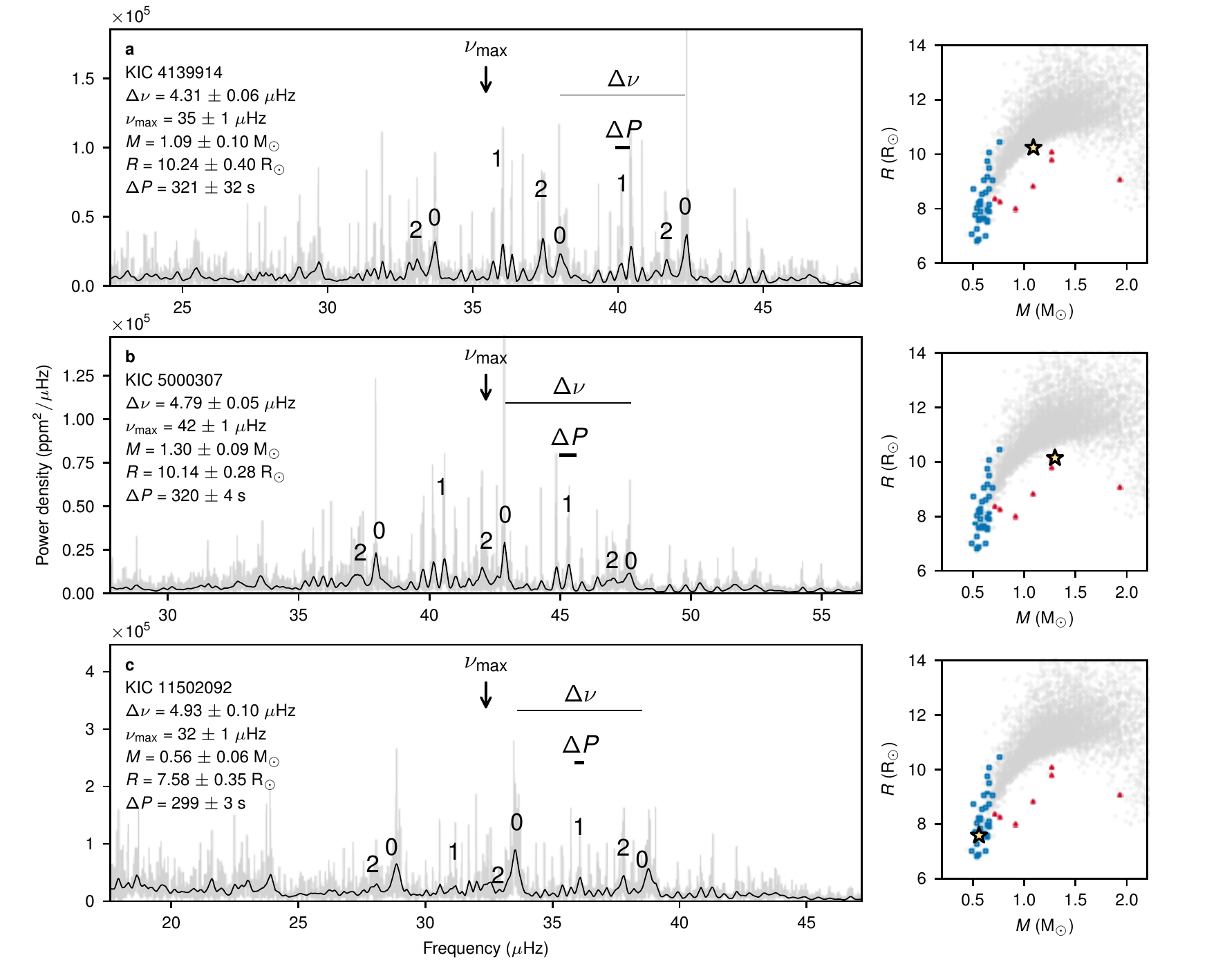}}
\caption{Power spectra for three representative stars, including a regular CHeB star (panel a), an underluminous star (panel b), and a very low-mass star (panel c). The right panels show their locations on the mass--radius diagram marked by the star symbols. The power spectra (grey lines) are smoothed by 0.06\Dnu{} (overlaid black lines). The integers $0$--$2$ represent the angular-degree $l$. The locations of \numax{} are indicated by the arrows. 
The observed values of \Dnu{} and \DP{} (see Extended Data Fig.~\ref{fig:echelle}) are represented by the lengths of the black line segments. \label{fig:spectra}}
\end{figure}

\begin{figure}
\centering
\showfigures{\includegraphics[width=\linewidth]{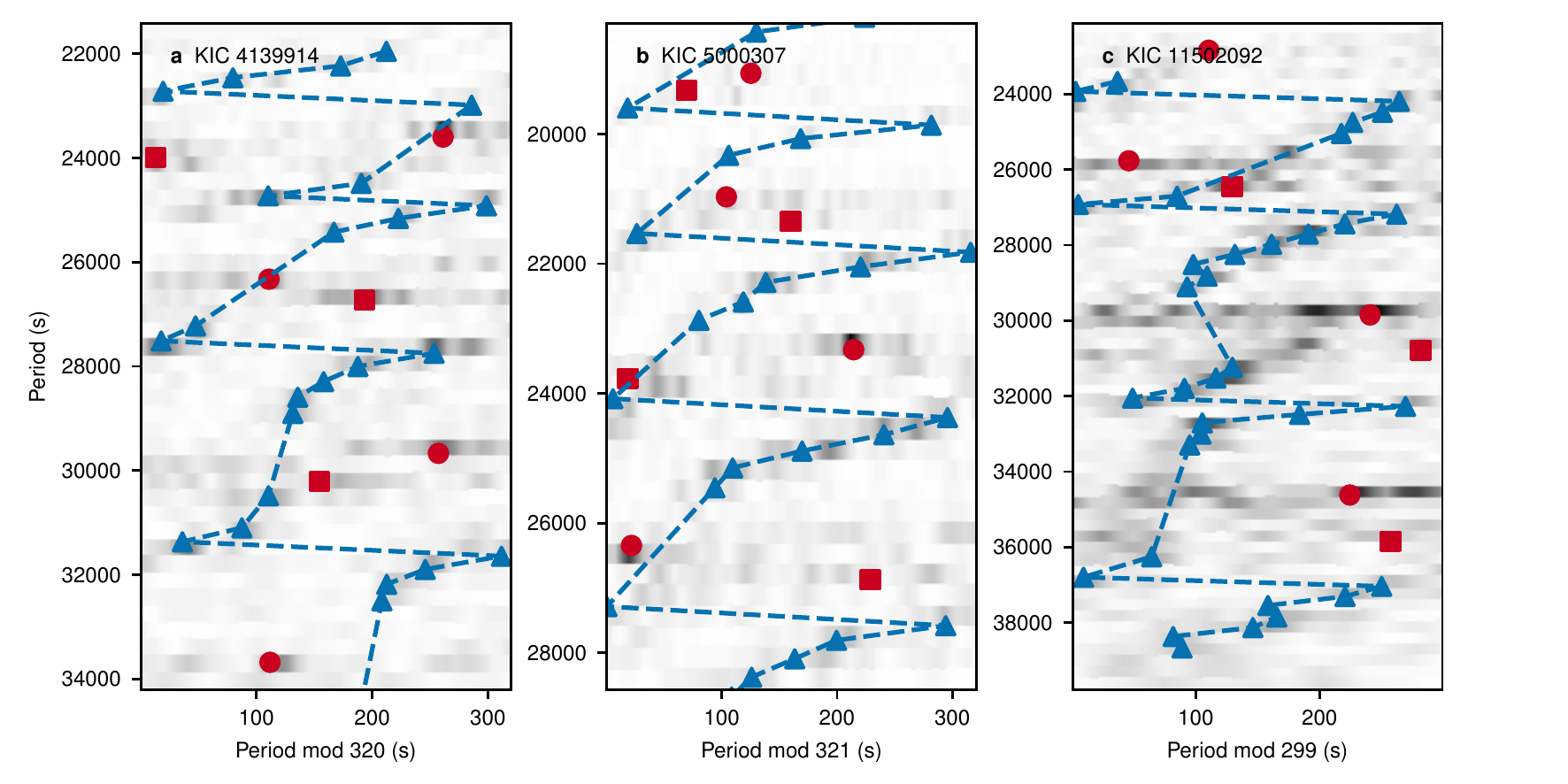}}
\caption{Period \'{e}chelle diagrams for the regular CHeB star (panel a), the underluminous star (panel b), and the very low-mass star (panel c) that are shown in Extended Fig.~\ref{fig:spectra}. The modes are marked by circles ($l=0$), triangles ($l=1$) and squares ($l=2$). Error bars are not shown. The blue dashed lines connect the $l=1$ modes in order. We adjusted the widths of the \'{e}chelle diagrams such that the $l=1$ modes form a ``zig-zag'' pattern\citep{bedding++2011-distinguish-rc-rgb}. Those widths correspond to the period spacings of $l=1$ modes, which confirm them as CHeB stars. \label{fig:echelle}}
\end{figure}

\begin{figure}
    \centering
    \showfigures{\includegraphics{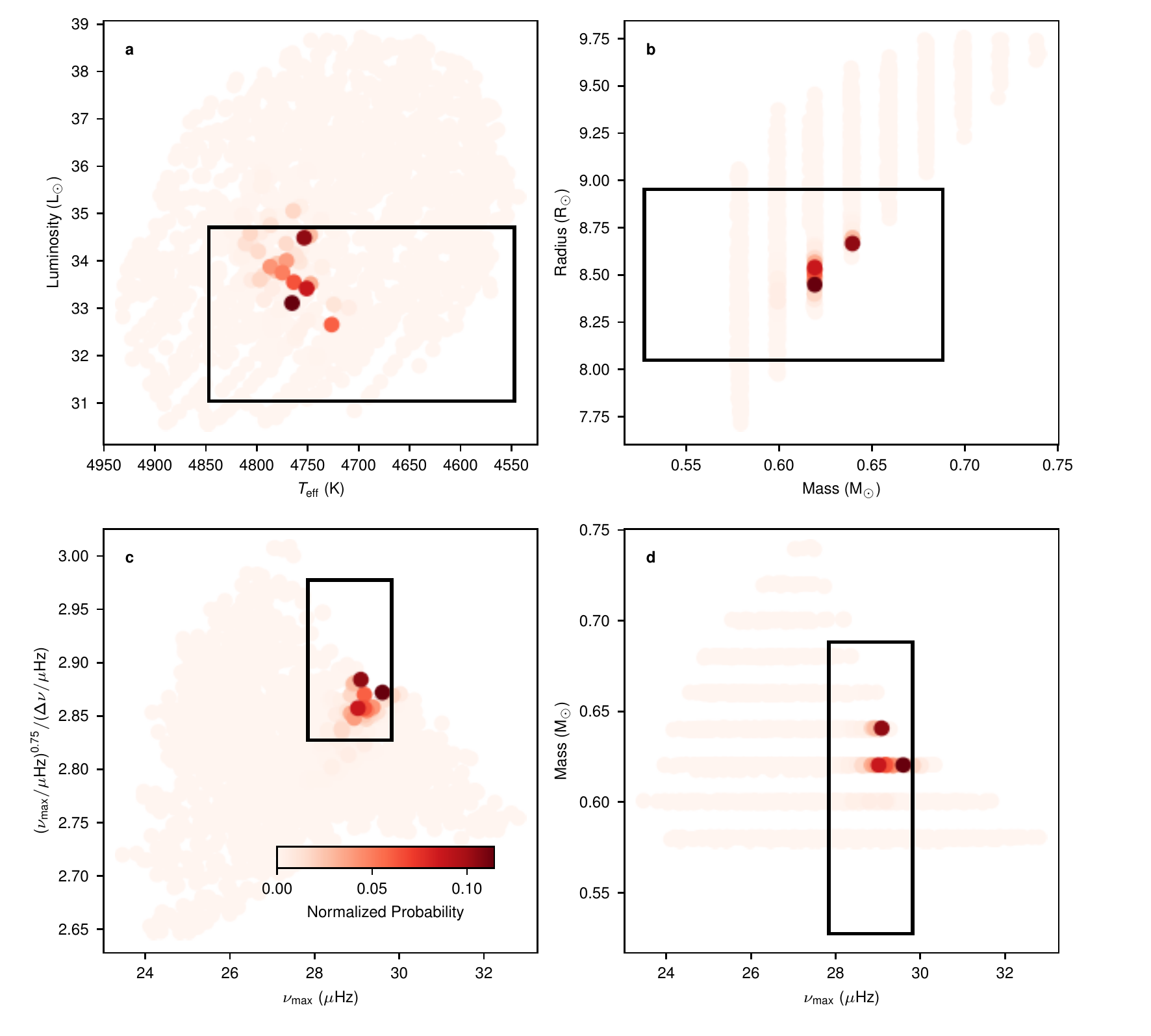}}
    \caption{Stellar models for KIC 8367834 within 3$\sigma$ of the classical constraints, colour-coded with probability using constraints from parallax, \Teff{}, metallicity, and oscillation frequencies. a): the H--R diagram; b): the mass--radius diagram; c): $\nu_{\rm max}^{0.75}/\Delta\nu$ vs. \numax{}; d): mass vs. \numax{}. The black boxes show the $1.5\sigma$ confidence regions, either directly from observations ($L$, \Teff{}, \numax{}, \Dnu{}) or from the scaling relations ($M$, $R$).\label{fig:sapphire}}
\end{figure}

\noindent\secondtitle{Identification of the underluminous stars} 
To set expectations of the sharpness of the ZAHeB edge, we followed our previous method described in Li et al.\citep{liyg++2021-sc-intrinsic-scatter}. This involved using a Galactic simulation sample generated by \galaxia{} \citep{Sharma++2011-galaxia,sharma++2019-k2-hermes-age-metallicity-thick-disc}, which has been tied to the \kepler{} target selection function.
In Extended Data Fig.~\ref{fig:galaxia}a, we show the \galaxia{} CHeB population on the $\nu_{\rm max}^{0.75}/\Delta\nu$--\Dnu{} diagram\citep{huber++2010-800-rg-kepler,bedding++2011-distinguish-rc-rgb,yuj++2018-16000-rg}.
Next, we identified the theoretical ZAHeB edge using a spline (the black dashed line) interpolated between several anchor points (the green crosses). 
We focused on the vertical distances to the edge because the horizontal direction has negligible uncertainties (in Fig.~\ref{fig:nike}b the errorbars on the red triangles are smaller than the symbol size). 
In the inset of Extended Data Fig.~\ref{fig:galaxia}a, we show the histogram of the vertical distances to the ZAHeB edge. 
Although the simulated sample forms a very sharp edge, it is still broadened by scatter in \Teff{} and [M/H].
To determine the intrinsic broadening, we fitted the distribution with a half-Gaussian half-Lorentzian profile. 
The intrinsic broadening $\sigma_{\rm intrinsic}$, measured by the standard deviation of the Gaussian profile, was $0.06$. 

\rev{We caution that the extremely metal-poor stars (e.g. [M/H]$=-2$ dex) could occupy the right side of the edge, although they are very rare in our sample. 
As shown in Fig.~\ref{fig:mass-feh}, most \kepler{} CHeB stars have [M/H]$ > -1.0$ dex. The identified underluminous stars (red triangles) have metallicities $> -0.5$ dex. This means that the right side of the edge is still a ``forbidden'' zone for these stars. }

Similarly, in Fig.\ref{fig:nike}b, we identified the observed ZAHeB edge for the \kepler{} sample with a spline and collected all stars that lay on the right of the edge. 
The statistical uncertainty was combined with $\sigma_{\rm intrinsic}$ in quadrature to represent the final uncertainty.
The underluminous stars were selected as being at least $1\sigma$ away from the observed ZAHeB edge in the vertical direction. 
\rev{We list the underluminous stars in the Supplementary Information. }

\noindent\secondtitle{Identification of the very low-mass stars} The lowest possible mass of a CHeB star, limited by the age of the universe (13.8\,Gyr\citep{planck++2015-primordial-helium}), is critically dependent on the metallicity.
Using stellar isochrones at 13.8 Gyr from MIST\citep{Choi++2016-mist-1-solar-scaled-models}, we extracted the model masses at helium-burning stage for different values of [M/H]. 
In Extended Data Fig.~\ref{fig:galaxia}b, we show the simulated \galaxia{} population on the [M/H]--mass diagram. The simulated sample forms a very sharp edge that coincides with the theoretical limit on mass (denoted by the dashed line).

The theoretical limit on mass (the solid line) is also shown in Fig.~\ref{fig:mass-feh}, together with the observed \kepler{} sample. We assume the mass loss driven by radiation and pulsation can at most lower this limit by $0.2$ \Msun{} (the dashed line).
Hence, we identified the stars at least $1\sigma$ to the left of the dashed line as the very low-mass stars, which must experience enhanced mass loss, possibly due to a companion.  
\rev{We list the very low-mass stars in  the Supplementary Information.}

\noindent\secondtitle{Modelling of a very low-mass star}
Since we rely on the scaling relations to derive stellar masses for the very low-mass stars, it is important to confirm the accuracy of the scaling relations in this regime. 
The \Dnu{} scaling relation can be checked with stellar models by calculating the mode frequencies and comparing with the density\citep{white++2011-asteroseismic-diagrams-cd-epsilon-deltaP-models,guggenberger++2016-metallicity-scaling-relation,sharma++2016-population-rg-kepler,rodrigues++2017-dpi-modelling,serenelli++2017-apokasc-dwarf-subgiant,pinsonneault++2018-apokasc}.
\rev{The models produce a correction factor which is applied to observations.}
The problem lies in the \numax{} scaling relation, which does not have a solid theoretical basis.
Zinn et al.\cite{zinn++2019-radius-sc} found no obvious difference between the scaling relation based radii and the Gaia radii for stars smaller than $R=30$ \Rsun{} within observational uncertainties.
Li et al.\cite{liyg++2021-sc-intrinsic-scatter} used the sharpness of the ZAHeB edge to conclude the \numax{} scaling relation has very small instinsic scatter of $~1.1\%$. 
However, the \numax{} scaling relation could perhaps have a systematic offset that bias stellar masses in this very low-mass regime.
In order to examine this, we used stellar modelling to show one of the very low-mass stars is indeed very low-mass, by constraining stellar models using luminosity, metallicity, \Teff{}, and oscillation mode frequencies. 
This does not use the information contained in the \numax{} scaling relation.

We chose the target KIC 8367834 because it has the best parallax among the very low-mass stars. 
We adopted the metallicity [M/H], $0.19\pm0.05$ dex, from APOGEE DR17\citep{abdurrouf++2021-apogee-dr17}.
We determined \Teff{} to be $4697\pm100$ K with the InfraRed Flux Method\citep{casagrande++2021-irfm-gaia}. 
Using ISOCLASSIFY\citep{huber++2017-seismic-radii-gaia,berger++2020-gaia-kepler-1-stars}, we derived trigonometric luminosities $L$, $32.87\pm1.22$~\Lsun{}, with the Gaia EDR3 parallax\citep{gaia-2016-mission,gaia-2020-edr3}, 2MASS J-band magnitudes, and extinctions from the dust map\cite{green++2019-dustmap}. 
We extracted 5 radial frequencies\citep{liyg++2020-kepler-36-subgiants}. They are $21.07\pm0.03$ $\mu$Hz, $25.31\pm0.02$ $\mu$Hz, $29.40\pm0.02$ $\mu$Hz, $33.89\pm0.02$ $\mu$Hz, and $38.25\pm0.03$ $\mu$Hz.

We constructed a grid of stellar models by varying metallicities [M/H] from $0.03$--$0.43$ dex in steps of $0.05$ dex ($Z$ from $0.0152$--$0.0357$), initial masses from $0.8$--$1.4$ \Msun{} in steps of $0.2$ \Msun{},  and final masses from $0.5$--$0.8$ \Msun{} in steps of $0.02$ \Msun{}. 
We first evolved models with various initial masses and metallicities until the onset of helium burning and saved these models.
These models then lost their outer envelopes at a rate of $10$ \Msun{}/Myr until the desired final masses were reached. 
The total mass loss spans shorter than 1\% of the helium burning lifetime.
The other model parameters were kept the same as the parameters we used to construct the models shown in Fig.~\ref{fig:nike}.
We calculated radial oscillation frequencies for all the models during the CHeB stage. 

We optimised the stellar models using a maximum likelihood approach:
\begin{equation}
    p \propto \exp(-\chi^2/2),
\end{equation}
where
\begin{equation}
    \chi^2 = \chi^2_{\rm classical} + \chi^2_{\rm seismic}.
\end{equation}
The classical constraints include three stellar properties, $q=\{ L, T_{\rm eff}, {\rm [M/H]} \}$:
\begin{equation} 
\chi^2_{\rm classical}  = \sum_q \frac{ \left[q_{{\rm mod}}-q_{{\rm obs}}\right]^2 }{ \sigma^2_{q} }.
\end{equation}
The seismic constraints include the extracted radial modes:
\begin{equation}\label{eq:chi2-seis}
\chi^2_{\rm seismic}  = \sum_n \frac{ \left[ \nu_{{\rm mod}, n}-\nu_{{\rm obs},n} \right]^2 }{ \sigma^2_{\nu_{{\rm mod}}} + \sigma^2_{\nu_{{\rm obs}, n}} }, \end{equation}
where $\sigma_{\nu_{{\rm mod}}}$ is a systematic uncertainty due to the limited resolution of the model grid \citep{litd++2020-kepler-36-subgiants,ong++2021-surf-corr-sg}. To evaluate $\sigma_{\nu_{{\rm mod}}}$, we first identified the best-fitting model (using Eq.~\ref{eq:chi2-seis} and treating $\sigma_{\nu_{{\rm mod}}}$ as 0) and calculated its root-mean-square difference. We also corrected the theoretical frequencies due to the surface effect with the inverse-cubic formula \citep{ball+2014-surface-correction-inertia-weighted}.

Extended Data Fig.~\ref{fig:sapphire} shows the stellar models within 3$\sigma$ of the classical constraints, colour-coded with the probability. 
Firstly, the most likely mass lies in the $1.5\sigma$ region determined from the scaling relations, validating the accuracy of the scaling relations. 
Secondly and unsurprisingly, the frequency modelling yields the stellar mass to a even greater precision, suggesting a more accurate method to determine masses.

\noindent\secondtitle{Rates of binary interactions}
We estimated the number of stars that would be expected to have lost mass due to binary interaction on the RGB. 
We used the observed masses and [M/H] of \kepler{} CHeB stars as the mass ($M_1$) and metallicity distributions of the primary stars in binary systems and calculated the maximum radius on the RGB, $R_{\rm RGB, max}$, with MIST stellar evolutionary tracks\citep{Choi++2016-mist-1-solar-scaled-models}.
Assuming circular orbits, we randomly sampled binary fractions $f$, orbital periods $P$ and mass ratios $q=M_1/M_2$ from observations of binary statistics\citep{moe+2017-p-q-binary} and derived the radius of the L1 Lagrangian point\citep{eggleton-1983-roche-lobes} according to
\begin{equation}
R_{\rm L1} = a \frac{0.49 q^{2/3}}{0.6q^{2/3} + \ln(1+q^{1/3})},
\end{equation}
where $a$ is the semi-major axis, which links to the orbital period,
\begin{equation}
P = 2\pi\sqrt{\frac{a^3}{G(M_1+M_2)}}.
\end{equation}
The stars that expand their envelopes on the RGB beyond the L1 Lagrangian point, i.e. $R_{\rm RGB, max}>R_{\rm L1}$, are subject to mass loss. The expected number of mass-loss stars is the sum of binary fractions $f$ for the stars which satisfy the above condition.

\noindent\secondtitle{Variations of radial velocities}
Using multi-epoch radial velocity data from APOGEE and LAMOST, we divided the maximal change $|{\rm RV}_{\rm max}-{\rm RV}_{\rm min}|$ by the median of statistical errors $e_{{\rm RV}}$ to represent RV variations\citep{mazzola-daher++2021-apogee-multiplicity} \rev{(see Supplementary Information)}.
\rev{Only a few of} the identified post-mass-transfer stars show large RV variations. 
To better understand the distributions, we generated RV time series ${\rm RV}_b$ at time $t$ for the simulated binary sample with
\begin{equation}
{\rm RV}_b(t) = 2\pi a /P \sin i^* \sin(2\pi t/P) + \epsilon_{{\rm RV}},  \end{equation}
where $i^*$ is the inclination angle drawn from an isotropic distribution, and $\epsilon_{{\rm RV}}$ is drawn from a normal distribution with standard deviation $e_{{\rm RV}}$. 
Both $t$ and $e_{{\rm RV}}$ used the observed RV 
measurements. 
Similarly, we generated RV time series for single stars with 
\begin{equation}
{\rm RV}_s(t) = \epsilon_{{\rm RV}}.
\end{equation}
Using the same method, we estimated the RV variations for the simulated samples. 
The comparison between \rev{the observed and the simulated samples} suggests that a significant RV variation ($|{\rm RV}_{\rm max}-{\rm RV}_{\rm min}|/e_{{\rm RV}} >10$) may indicate a binary system, but a small RV variation does not necessarily exclude binarity. 
\rev{This explains the lack of RV variations in most stars identified in this work.}

\section*{Data Availability}
We made use of publicly available data in this work.
\kepler{} data are available from the MAST portal at {https://archive.stsci.edu/access-mast-data}, 
APOGEE data at {https://www.sdss.org/dr16/},
LAMOST data at {http://dr5.lamost.org/v3/doc/vac} and {https://github.com/hypergravity/paperdata}, and
\gaia{} data at {https://gea.esac.esa.int/archive/}.
The data needed to reproduce this work is available at GitHub (https://github.com/parallelpro/Yaguang\_stripped\_rg\_repo). All other data are available from the corresponding author upon reasonable request.

\section*{Code Availability}
This work is made possible by the following open-source software: {Numpy}\citep{numpy}, {Scipy}\citep{scipy}, {Matplotlib}\citep{matplotlib}, {Astropy}\citep{astropy1,astropy2}, {Pandas}\citep{pandas}, {MESA}\citep{paxton++2011-mesa,paxton++2013-mesa,paxton++2015-mesa,paxton++2018-mesa,paxton++2019-mesa}, {MESASDK}\citep{mesasdk}, {GYRE}\citep{townsend+2013-gyre}, {pySYD}\citep{chontos++2021-pysyd}, {Lightkurve}\citep{lightkurve}, {EchellePlotter} (https://github.com/9yifanchen9/EchellePy), {ISOCLASSIFY}\citep{huber++2017-seismic-radii-gaia,berger++2020-gaia-kepler-1-stars}. 
The scripts used in this work is available at a curated GitHub repository (https://github.com/parallelpro/Yaguang\_stripped\_rg\_repo).

\section*{Acknowledgements}
We thank Mark Hon, Karsten Brogaard, and Yvonne Elsworth for their comments. 

T.R.B and D.H. acknowledge funding from the Australian Research Council (Discovery Project DP210103119). D.H. also acknowledges support from the Alfred P. Sloan Foundation and the National Aeronautics and Space Administration (80NSSC19K0597). M.J. acknowledges the Lasker Fellowship grant. S.B. acknowledges the Joint Research Fund in Astronomy (U2031203) under cooperative agreement between the National Natural Science Foundation of China (NSFC) and Chinese Academy of Sciences (CAS) and the NSFC grants 12090040 and 12090042. G.L. acknowledges support from the project BEAMING ANR-18-CE31-0001 of the French National Research Agency (ANR) and from the Centre National d’Etudes Spatiales (CNES).

We gratefully acknowledge the Kepler teams, whose efforts made these results possible.
Funding for the Kepler mission is provided by the NASA Science Mission Directorate. This paper includes data collected by the Kepler mission and obtained from the MAST data archive at the Space Telescope Science Institute (STScI). STScI is operated by the Association of Universities for Research in Astronomy, Inc., under NASA contract NAS 5–26555.

Guoshoujing Telescope (the Large Sky Area Multi-Object Fiber Spectroscopic Telescope LAMOST) is a National Major Scientific Project built by the Chinese Academy of Sciences. Funding for the project has been provided by the National Development and Reform Commission. LAMOST is operated and managed by the National Astronomical Observatories, Chinese Academy of Sciences.

This work presents results from the European Space Agency (ESA) space mission Gaia. Gaia data are being processed by the Gaia Data Processing and Analysis Consortium (DPAC). Funding for the DPAC is provided by national institutions, in particular the institutions participating in the Gaia MultiLateral Agreement (MLA). The Gaia mission website is https://www.cosmos.esa.int/gaia. The Gaia archive website is https://archives.esac.esa.int/gaia.

Funding for the Sloan Digital Sky Survey IV has been provided by the Alfred P. Sloan Foundation, the U.S. Department of Energy Office of Science, and the Participating Institutions. 

We acknowledge the Sydney Informatics (a core research facility of the University of Sydney), 
high performance computing (HPC) cluster Artemis from the University of Sydney, 
HPC cluster headnode from the School of Physics,
and HPC cluster gadi from the National Computational Infrastructure (NCI Australia, an NCRIS enabled capability supported by the Australian Government),
for providing the HPC resources that have contributed to the research results reported within this paper.


\section*{Author Contributions}
Y.L., T.R.B., D.S., Y.C., I.L.C. and G.L. analysed photometric data;
S.J.M, D.H., X.Z., S.B and D.R.H. contributed to binary confirmation;
Y.L., M.J. and D.M. constructed theoretical models;
B.T.M, M.R.H, S.S and Y.W. interpreted spectroscopic data.
All authors discussed the results and commented on the manuscript.


\section*{Competing Interests}
The authors declare no competing interests.

\section*{Correspondence}
Correspondence should be addressed to Y.L. (yaguang.li@sydney.edu.au) or T.R.B. (tim.bedding@sydney.edu.au).


\section*{Figure Captions}

\bigskip

\bigskip

\bigskip

\bigskip
\extended

\section*{Extended Data Figure Captions}

\bigskip

\bigskip

\bigskip

\bigskip

\bigskip


\bigskip
\clearpage

\supplementary

\begin{figure}
\centering
\includegraphics[width=88mm]{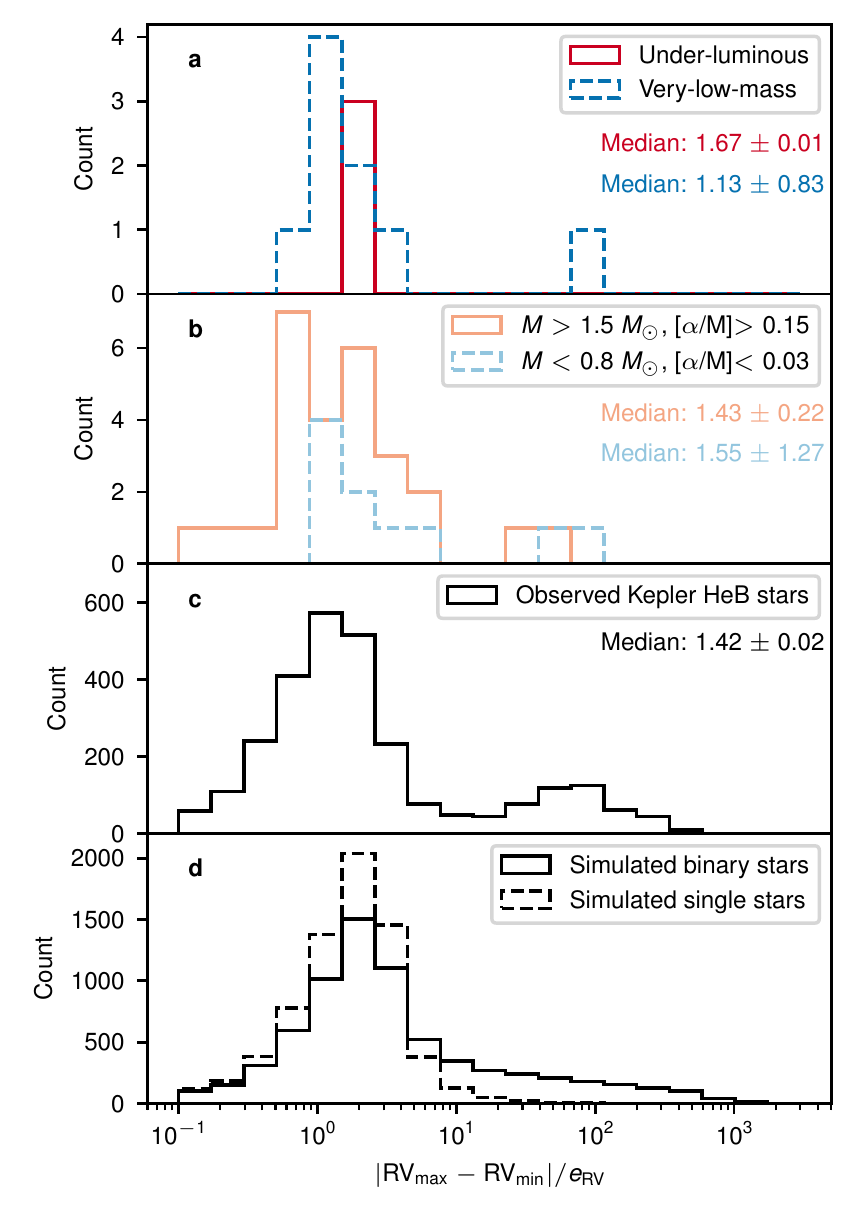}\\
\caption{Distributions of maximal change in radial velocity (RV) scaled by statistical errors, for a) the under-luminous and low-mass stars identified from Figs. 1 and 2, b) the high-mass $\alpha$-rich stars and low-mass low-$\alpha$ stars identified from Fig. 3, c) the \kepler{} CHeB red giant sample, and d) the simulated binary and single stars (see Methods). \label{fig:rv}}
\end{figure}

\clearpage

\begin{table}\footnotesize
\caption{Stellar parameters of the under-luminous stars. 
$\Delta\nu$ is the p-mode large seperation, determined using radial mode frequencies; $\nu_{\rm max}$, the frequency of maximum power \citep{yuj++2018-16000-rg}; $T_{\rm eff}$, effective temperature \citep{mathur++2017-revised-properties-kepler-targets-dr25,yuj++2018-16000-rg}; [M/H], metallicity; Ref, the reference for metallicity (a: \citep{abdurrouf++2021-apogee-dr17}; b: \citep{xiang++2019-lamost-dr5}; c: \citep{mathur++2017-revised-properties-kepler-targets-dr25,yuj++2018-16000-rg}); mass, stellar mass estimated with the scaling relations; radius, stellar radius estimated with the scaling relations; $d/\sigma$, the vertical distance to the ZAHeB edge scaled by uncertainty (see Methods and Fig. 1).}\\ 
\begin{tabular*}{\columnwidth}{@{\extracolsep{\fill}}lrrrrrrrrrrr}
\toprule
     KIC & $\Delta\nu$ ($\mu$Hz) & $\nu_{\rm max}$ ($\mu$Hz) &  $T_{\rm eff}$ (K) &        [M/H] (dex) & Ref &    Mass (M$_\odot$) &   Radius (R$_\odot$) & $d/\sigma$ \\
\midrule
 3963011 &   $4.82$ $\pm$ $0.06$ &        $34.2$ $\pm$ $0.9$ & $5099$ $\pm$ $151$ & -0.54 $\pm$ $0.30$ & $c$ & $0.73$ $\pm$ $0.08$ &  $8.37$ $\pm$ $0.33$ &     $1.55$ \\
 4755614 &   $5.00$ $\pm$ $0.05$ &        $44.2$ $\pm$ $0.5$ & $5102$ $\pm$ $165$ & -0.01 $\pm$ $0.10$ & $a$ & $1.30$ $\pm$ $0.09$ &  $9.85$ $\pm$ $0.28$ &     $3.60$ \\
 5000307 &   $4.79$ $\pm$ $0.05$ &        $42.2$ $\pm$ $0.6$ &  $4955$ $\pm$ $80$ & -0.28 $\pm$ $0.10$ & $a$ & $1.30$ $\pm$ $0.09$ & $10.14$ $\pm$ $0.28$ &     $2.08$ \\
 8145590 &   $5.71$ $\pm$ $0.07$ &        $49.5$ $\pm$ $0.4$ &  $4807$ $\pm$ $80$ & -0.05 $\pm$ $0.10$ & $a$ & $0.98$ $\pm$ $0.06$ &  $8.20$ $\pm$ $0.22$ &     $8.03$ \\
 8489112 &   $7.00$ $\pm$ $0.05$ &        $76.0$ $\pm$ $1.0$ & $5389$ $\pm$ $162$ & -0.52 $\pm$ $0.10$ & $b$ & $1.93$ $\pm$ $0.13$ &  $9.00$ $\pm$ $0.22$ &     $2.58$ \\
10665157 &   $5.14$ $\pm$ $0.09$ &        $36.2$ $\pm$ $0.5$ & $5475$ $\pm$ $164$ & -0.47 $\pm$ $0.30$ & $c$ & $0.73$ $\pm$ $0.07$ &  $8.01$ $\pm$ $0.33$ &     $1.62$ \\
10724735 &   $5.34$ $\pm$ $0.07$ &        $48.2$ $\pm$ $0.6$ &  $4757$ $\pm$ $80$ &  0.14 $\pm$ $0.10$ & $a$ & $1.17$ $\pm$ $0.08$ &  $9.08$ $\pm$ $0.26$ &     $4.23$ \\
\bottomrule
\end{tabular*}
\end{table}

\clearpage

\begin{table}\footnotesize
\caption{Stellar parameters of the very-low-mass stars. 
$\Delta\nu$ is the p-mode large seperation, determined using radial mode frequencies; $\nu_{\rm max}$, the frequency of maximum power \citep{yuj++2018-16000-rg}; $T_{\rm eff}$, effective temperature \citep{mathur++2017-revised-properties-kepler-targets-dr25,yuj++2018-16000-rg}; [M/H], metallicity; Ref, the reference for metallicity (a: \citep{abdurrouf++2021-apogee-dr17}; b: \citep{xiang++2019-lamost-dr5}); mass, stellar mass estimated with the scaling relations; radius, stellar radius estimated with the scaling relations; $d/\sigma$, the horizontal distance to the lower limit on mass scaled by uncertainty (see Methods and Fig. 2).}\\
\begin{tabular*}{\columnwidth}{@{\extracolsep{\fill}}lrrrrrrrrr}
\toprule
     KIC & $\Delta\nu$ ($\mu$Hz) & $\nu_{\rm max}$ ($\mu$Hz) &  $T_{\rm eff}$ (K) &        [M/H] (dex) & Ref &    Mass (M$_\odot$) &   Radius (R$_\odot$) & $d/\sigma$ \\
\midrule
 2285898 &   $3.83$ $\pm$ $0.08$ &        $24.8$ $\pm$ $0.6$ &  $4774$ $\pm$ $80$ & -0.23 $\pm$ $0.10$ & $a$ & $0.62$ $\pm$ $0.07$ &  $9.22$ $\pm$ $0.44$ &     $1.19$ \\
 2305082 &   $4.27$ $\pm$ $0.05$ &        $27.9$ $\pm$ $0.4$ & $5055$ $\pm$ $163$ & -0.41 $\pm$ $0.10$ & $b$ & $0.61$ $\pm$ $0.05$ &  $8.48$ $\pm$ $0.28$ &     $1.20$ \\
 2861630 &   $4.24$ $\pm$ $0.04$ &        $28.8$ $\pm$ $1.0$ &  $4839$ $\pm$ $80$ &  0.12 $\pm$ $0.10$ & $a$ & $0.67$ $\pm$ $0.08$ &  $8.83$ $\pm$ $0.38$ &     $1.45$ \\
 3335197 &   $5.32$ $\pm$ $0.09$ &        $35.6$ $\pm$ $0.6$ & $5338$ $\pm$ $164$ & -0.11 $\pm$ $0.10$ & $b$ & $0.60$ $\pm$ $0.06$ &  $7.34$ $\pm$ $0.29$ &     $2.16$ \\
 3534438 &   $5.05$ $\pm$ $0.07$ &        $33.4$ $\pm$ $1.1$ & $4755$ $\pm$ $166$ & -0.06 $\pm$ $0.10$ & $b$ & $0.52$ $\pm$ $0.07$ &  $7.29$ $\pm$ $0.35$ &     $3.09$ \\
 3839443 &   $5.86$ $\pm$ $0.02$ &        $39.1$ $\pm$ $0.6$ & $5143$ $\pm$ $147$ & -0.12 $\pm$ $0.10$ & $a$ & $0.52$ $\pm$ $0.03$ &  $6.60$ $\pm$ $0.14$ &     $4.92$ \\
 4813529 &   $5.74$ $\pm$ $0.08$ &        $37.1$ $\pm$ $0.5$ & $5243$ $\pm$ $150$ & -0.24 $\pm$ $0.10$ & $a$ & $0.49$ $\pm$ $0.04$ &  $6.54$ $\pm$ $0.22$ &     $4.48$ \\
 5271626 &   $3.86$ $\pm$ $0.06$ &        $25.1$ $\pm$ $0.5$ &  $4780$ $\pm$ $80$ &  0.03 $\pm$ $0.10$ & $a$ & $0.63$ $\pm$ $0.05$ &  $9.27$ $\pm$ $0.34$ &     $2.12$ \\
 5388291 &   $4.46$ $\pm$ $0.21$ &        $28.8$ $\pm$ $1.3$ & $4723$ $\pm$ $146$ & -0.01 $\pm$ $0.10$ & $b$ & $0.54$ $\pm$ $0.13$ &  $8.03$ $\pm$ $0.85$ &     $1.60$ \\
 5437901 &   $5.09$ $\pm$ $0.16$ &        $32.9$ $\pm$ $0.6$ &  $5123$ $\pm$ $80$ & -0.12 $\pm$ $0.10$ & $b$ & $0.53$ $\pm$ $0.07$ &  $7.31$ $\pm$ $0.48$ &     $2.49$ \\
 5522844 &   $4.72$ $\pm$ $0.08$ &        $32.6$ $\pm$ $0.7$ & $5103$ $\pm$ $156$ &  0.05 $\pm$ $0.10$ & $a$ & $0.68$ $\pm$ $0.07$ &  $8.28$ $\pm$ $0.36$ &     $1.19$ \\
 6790379 &   $4.39$ $\pm$ $0.03$ &        $28.6$ $\pm$ $0.6$ & $4845$ $\pm$ $148$ & -0.46 $\pm$ $0.10$ & $b$ & $0.56$ $\pm$ $0.04$ &  $8.10$ $\pm$ $0.24$ &     $2.13$ \\
 6964709 &   $5.15$ $\pm$ $0.20$ &        $32.9$ $\pm$ $0.6$ & $5286$ $\pm$ $173$ & -0.38 $\pm$ $0.10$ & $a$ & $0.50$ $\pm$ $0.08$ &  $6.99$ $\pm$ $0.56$ &     $2.06$ \\
 7039616 &   $4.26$ $\pm$ $0.10$ &        $27.3$ $\pm$ $0.3$ & $4816$ $\pm$ $168$ & -0.16 $\pm$ $0.10$ & $b$ & $0.56$ $\pm$ $0.06$ &  $8.34$ $\pm$ $0.42$ &     $2.36$ \\
 7440589 &   $4.16$ $\pm$ $0.05$ &        $29.4$ $\pm$ $0.4$ &  $4754$ $\pm$ $80$ &  0.15 $\pm$ $0.10$ & $a$ & $0.73$ $\pm$ $0.05$ &  $9.19$ $\pm$ $0.27$ &     $1.15$ \\
 7522091 &   $3.56$ $\pm$ $0.02$ &        $24.4$ $\pm$ $0.4$ &  $4563$ $\pm$ $80$ &  0.17 $\pm$ $0.10$ & $a$ & $0.73$ $\pm$ $0.04$ & $10.20$ $\pm$ $0.20$ &     $1.36$ \\
 8019232 &   $4.94$ $\pm$ $0.12$ &        $32.2$ $\pm$ $0.5$ & $5222$ $\pm$ $150$ & -0.21 $\pm$ $0.10$ & $a$ & $0.57$ $\pm$ $0.07$ &  $7.61$ $\pm$ $0.40$ &     $1.91$ \\
 8245132 &   $5.16$ $\pm$ $0.11$ &        $33.7$ $\pm$ $0.7$ & $5366$ $\pm$ $170$ & -0.10 $\pm$ $0.10$ & $a$ & $0.58$ $\pm$ $0.06$ &  $7.41$ $\pm$ $0.36$ &     $2.25$ \\
 8299794 &   $5.90$ $\pm$ $0.07$ &        $40.8$ $\pm$ $0.8$ & $5367$ $\pm$ $161$ & -0.28 $\pm$ $0.10$ & $b$ & $0.58$ $\pm$ $0.05$ &  $6.77$ $\pm$ $0.24$ &     $1.93$ \\
 8367834 &   $4.29$ $\pm$ $0.06$ &        $28.8$ $\pm$ $0.7$ &  $4723$ $\pm$ $80$ &  0.19 $\pm$ $0.10$ & $a$ & $0.61$ $\pm$ $0.06$ &  $8.52$ $\pm$ $0.31$ &     $3.06$ \\
 8669094 &   $3.59$ $\pm$ $0.06$ &        $22.6$ $\pm$ $0.7$ & $4847$ $\pm$ $100$ & -0.20 $\pm$ $0.10$ & $b$ & $0.63$ $\pm$ $0.07$ &  $9.69$ $\pm$ $0.46$ &     $1.06$ \\
 9330853 &   $5.28$ $\pm$ $0.01$ &        $35.5$ $\pm$ $0.5$ & $5487$ $\pm$ $164$ & -0.34 $\pm$ $0.10$ & $b$ & $0.61$ $\pm$ $0.04$ &  $7.36$ $\pm$ $0.15$ &     $1.73$ \\
 9644558 &   $4.41$ $\pm$ $0.06$ &        $28.3$ $\pm$ $0.7$ & $5099$ $\pm$ $100$ & -0.44 $\pm$ $0.10$ & $a$ & $0.57$ $\pm$ $0.06$ &  $8.17$ $\pm$ $0.32$ &     $1.52$ \\
 9783656 &   $5.03$ $\pm$ $0.02$ &        $33.9$ $\pm$ $0.4$ & $5331$ $\pm$ $169$ & -0.23 $\pm$ $0.10$ & $a$ & $0.63$ $\pm$ $0.04$ &  $7.73$ $\pm$ $0.17$ &     $1.62$ \\
 9814943 &   $4.69$ $\pm$ $0.05$ &        $30.8$ $\pm$ $0.5$ &  $4997$ $\pm$ $80$ & -0.16 $\pm$ $0.10$ & $a$ & $0.58$ $\pm$ $0.04$ &  $7.90$ $\pm$ $0.22$ &     $2.99$ \\
 9847893 &   $3.76$ $\pm$ $0.02$ &        $22.2$ $\pm$ $0.4$ &  $4905$ $\pm$ $80$ & -0.48 $\pm$ $0.10$ & $a$ & $0.51$ $\pm$ $0.03$ &  $8.74$ $\pm$ $0.19$ &     $3.77$ \\
10518222 &   $5.04$ $\pm$ $0.09$ &        $34.4$ $\pm$ $0.7$ &  $4995$ $\pm$ $80$ & -0.03 $\pm$ $0.10$ & $a$ & $0.60$ $\pm$ $0.06$ &  $7.64$ $\pm$ $0.31$ &     $2.37$ \\
10860146 &   $3.52$ $\pm$ $0.05$ &        $22.8$ $\pm$ $0.8$ &  $4667$ $\pm$ $80$ &  0.15 $\pm$ $0.10$ & $a$ & $0.65$ $\pm$ $0.08$ &  $9.92$ $\pm$ $0.45$ &     $1.63$ \\
11408704 &   $3.95$ $\pm$ $0.04$ &        $26.1$ $\pm$ $0.5$ &  $4702$ $\pm$ $80$ &  0.04 $\pm$ $0.10$ & $a$ & $0.63$ $\pm$ $0.05$ &  $9.12$ $\pm$ $0.28$ &     $2.38$ \\
11450315 &   $5.00$ $\pm$ $0.07$ &        $32.9$ $\pm$ $0.4$ &  $4950$ $\pm$ $80$ &  0.02 $\pm$ $0.10$ & $a$ & $0.55$ $\pm$ $0.04$ &  $7.45$ $\pm$ $0.23$ &     $4.52$ \\
11502092 &   $4.93$ $\pm$ $0.10$ &        $32.4$ $\pm$ $0.6$ &  $5016$ $\pm$ $80$ & -0.06 $\pm$ $0.10$ & $a$ & $0.56$ $\pm$ $0.06$ &  $7.58$ $\pm$ $0.35$ &     $2.89$ \\
12505644 &   $4.62$ $\pm$ $0.08$ &        $27.9$ $\pm$ $0.5$ & $5444$ $\pm$ $161$ & -0.48 $\pm$ $0.10$ & $b$ & $0.52$ $\pm$ $0.05$ &  $7.71$ $\pm$ $0.32$ &     $2.42$ \\
\bottomrule
\end{tabular*}
\end{table}


\clearpage

\firsttitle{References}
\bigskip\bigskip\bigskip


\clearpage

\end{document}